\documentclass[aps,prl,superscriptaddress,twocolumn,showpacs,preprintnumbers,amsmath,amssymb]{revtex4}
\usepackage[dvips]{graphicx,color}% Include figure files
\usepackage{dcolumn}% Align table columns on decimal point
\usepackage{bm}% bold math
\usepackage{ulem}
\usepackage{soul}
\begin{document}

\title{Experimental Verification of Comparability between Spin-Orbit and Spin-Diffusion Lengths}

\author{Yasuhiro Niimi}
\email{niimi@issp.u-tokyo.ac.jp}
\affiliation{Institute for Solid State Physics, University of Tokyo, 5-1-5 Kashiwa-no-ha, Kashiwa, Chiba 277-8581, Japan}
\author{Dahai Wei}
\affiliation{Institute for Solid State Physics, University of Tokyo, 5-1-5 Kashiwa-no-ha, Kashiwa, Chiba 277-8581, Japan}
\author{Hiroshi Idzuchi}
\affiliation{Institute for Solid State Physics, University of Tokyo, 5-1-5 Kashiwa-no-ha, Kashiwa, Chiba 277-8581, Japan}
\author{Taro Wakamura}
\affiliation{Institute for Solid State Physics, University of Tokyo, 5-1-5 Kashiwa-no-ha, Kashiwa, Chiba 277-8581, Japan}
\author{Takeo Kato}
\affiliation{Institute for Solid State Physics, University of Tokyo, 5-1-5 Kashiwa-no-ha, Kashiwa, Chiba 277-8581, Japan}
\author{YoshiChika Otani}
\affiliation{Institute for Solid State Physics, University of Tokyo, 5-1-5 Kashiwa-no-ha, Kashiwa, Chiba 277-8581, Japan}
\affiliation{RIKEN-ASI, 2-1 Hirosawa, Wako, Saitama 351-0198, Japan}

\date{January 3, 2013}

\begin{abstract}
We experimentally confirmed that the spin-orbit lengths 
of noble metals obtained from weak anti-localization measurements 
are comparable to the spin diffusion lengths determined 
from lateral spin valve ones. 
Even for metals with strong spin-orbit interactions such as Pt, 
we verified that the two methods gave comparable values 
which were much larger than those obtained from recent spin torque 
ferromagnetic resonance measurements. 
To give a further evidence for the comparability between 
the two length scales, 
we measured the disorder dependence of 
the spin-orbit length of copper by changing the thickness of the wire. 
The obtained spin-orbit length nicely 
follows a linear law as a function of the diffusion coefficient, 
clearly indicating that the Elliott-Yafet mechanism is dominant 
as in the case of the spin diffusion length.
\end{abstract}


\maketitle

Spin relaxation and spin dephasing are the central issues 
in the field of spintronics as they determine how long and far 
electrons can transfer 
the spin information~\cite{zutic_rmp_2004,fabian_review_1999}. 
Owing to recent technological advancements~\cite{yang_nphys_2008,fukuma_apl_2010,fukuma_nmat_2011,takahashi_apl_2012,seki_nmat_2008,niimi_prl_2011,niimi_prl_2012,liu_prl_2011,liu_science_2012,liu_prl_2012}, 
one can create the spin accumulation, i.e., 
the electrochemical potential difference 
between spin-up and down electrons at the Fermi level, 
$10\sim100$ times larger than that generated 
in conventional lateral spin valve devices or 
along edges of samples with spin Hall effects (SHEs). 
Such a large spin accumulation can induce a large pure spin current, 
the flow of spin angular momentum 
with no net charge current~\cite{takahashi_review_2008}. 
As its magnitude scales with the spin relaxation length 
or the spin diffusion length (SDL), the
quantitative evaluation of the SDL is of importance. 

Recent reports on magnetization switching at very thin ferromagnet/nonmagnet 
bilayer films~\cite{liu_science_2012,liu_prl_2012,miron_nature_2011} 
have triggered a heavy debate on the detailed mechanism. 
The first realization of the switching was reported by 
Miron~\textit{et al}.~\cite{miron_nature_2011} who 
concluded that the magnetization switching originates from the Rashba effect 
at the ferromagnet(Co)/nonmagnet(Pt) interface. 
Similar measurements were also performed by Liu~\textit{et al}. with a 
Co/Pt bilayer film~\cite{liu_prl_2012} as well as with 
a CoFeB/Ta bilayer one~\cite{liu_science_2012}. 
They claimed that the switching is due to the perpendicularly induced 
spin currents via the SHE of Pt and Ta. 
To discuss the conversion efficiency from charge current 
to spin current i.e., the spin Hall (SH) angle, in their devices, 
they performed the spin torque induced ferromagnetic resosnance (FMR) 
measurements and estimated the SH angle of Pt and Ta 
to be 0.07 and 0.15, 
respectively~\cite{liu_prl_2011,liu_prl_2012,liu_science_2012}.
These SH angles, however, are quite different from those obtained 
from the spin absorption method 
(0.02 for Pt and 0.004 fo Ta)~\cite{morota_prb_2011}. 
Furthermore Liu~\textit{et al}.~\cite{liu_arXiv_2011} pointed out that 
the overestimation of the SDLs reported in Ref.~\cite{morota_prb_2011} 
results in a large underestimation of the SH angles. 
To settle down such heavy debates, one needs another reliable way to 
estimate the SDL or the SH angle. 
In this Letter, we focus on weak anti-localization (WAL) observed 
in nonmagnetic metals. We first demonstrate 
that the spin-orbit (SO) length $L_{\rm SO}$ obtained from the WAL curve of Ag
is comparable to the SDL $L_{\rm s}$ estimated from the lateral spin valve 
measurement. 
We then extend a similar discussion to a strong SO material such as Pt. 
In metallic systems where the elastic mean free path $l_{e}$ is 
much shorter than $L_{\rm s}$, the dominant spin relaxation process is 
the Eliott-Yafet (EY) mechanism~\cite{elliott_yafet}. 
We also confirm that the EY mechanism works very well for $L_{\rm SO}$ by 
changing the diffusion coefficient $D$ of Cu wires. This experimental fact 
also verifies the comparability between $L_{\rm SO}$ and $L_{\rm s}$.

We prepared two types of devices, i.e., 
samples for WAL measurements and those
for spin injection measurements. 
Both samples were fabricated on a thermally-oxidized silicon substrate 
using electron beam lithography on polymethyl-methacrylate resist 
and a subsequent lift-off process. 
For the WAL samples, we prepared $\sim1$~mm long and 100 nm wide Ag (99.99\%), 
Cu (99.9999\%), and Pt (99.98\%) wires and 
performed the standard 4-probe measurement using a $^{3}$He cryostat. 
In order to obtain a very small WAL signal compared 
to the background resistance, we used a bridge circuit~\cite{niimi_prb_2010}. 
For the spin injection (or spin valve) measurement, 
we first prepared two Permalloy (Ni$_{81}$Fe$_{19}$; hereafter Py) wires, 
which work as spin injector and detector. 
To measure $L_{\rm s}$ of a strong SO material such as Pt, 
we inserted it in between the two Py wires. 
The three wires were bridged by a thicker Cu wire 
to transfer a pure spin current 
generated at the Py/Cu interface. 
%The sample demensions used for the spin valve measurements are the 
%same as in Refs.~\cite{niimi_prl_2011,niimi_prl_2012}.
To check the reproducibility, we measured at least a few different samples on 
the same batch both for the WAL and spin valve measurements.

\begin{figure}
\begin{center}
\includegraphics[width=5.5cm]{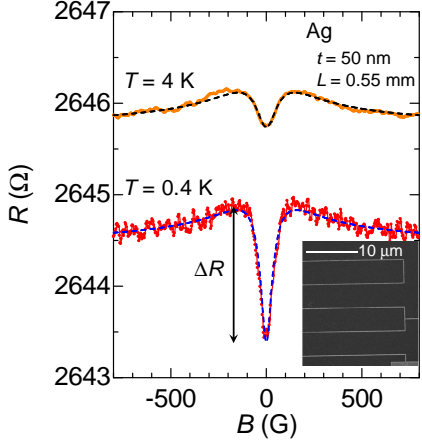}
\caption{(Color online) WAL curves of Ag wire measured at $T=0.4$ and 4~K. The broken lines are the best fits of Eq.~(\ref{eq1}). The inset shows a scanning electron micrograph of the Ag wire.} \label{fig1}
\end{center}
\end{figure}

In order to clarify whether $L_{\rm SO}$ from WAL measurements is 
equivalent to $L_{\rm s}$ from spin valve ones, 
we first measure WAL curves of a weak SO material. 
Figure~1 shows typical WAL curves of a Ag wire 
measured at $T = 0.4$ and 4~K. 
Unlike a normal weak localization (WL) curve, 
the resistance increases with 
increasing the perpendicular magnetic field $B$
because of the SO interaction. 
With decreasing temperature, the phase coherence of electrons 
gets larger and the WAL peak also gets sharper. 
The WAL peak of quasi one-dimensional (1D) wire 
can be fitted by the Hikami-Larkin-Nagaoka formula~\cite{gilles_book};
\begin{eqnarray}
\frac{\Delta R}{R_{\infty}} = \frac{1}{\pi L}\frac{R_{\infty}}{\hbar/e^{2}}
\left( \frac{\frac{3}{2}}{\sqrt{\frac{1}{L_{\varphi}^{2}}+\frac{4}{3}\frac{1}{L_{\rm SO}^{2}}+\frac{1}{3}\frac{w^{2}}{l_{B}^{4}}}} 
- \frac{\frac{1}{2}}{\sqrt{\frac{1}{L_{\varphi}^{2}} + \frac{1}{3}\frac{w^{2}}{l_{B}^{4}}}} \right) \label{eq1}
\end{eqnarray}
where $\Delta R$, $R_{\infty}$, $L$, and $w$ are respectively 
the WL correction factor, the resistance of the wire at high enough field, 
the length and width of the quasi-1D wire. 
$e$, $\hbar$, and $l_{B} = \sqrt{\hbar/eB}$ are respectively 
the electron charge, the reduced Plank constant, and 
the magnetic length. 
In Eq.~(\ref{eq1}), we have only two unknown parameters; 
$L_{\varphi}$ and $L_{\rm SO}$. 
According to the Fermi liquid theory~\cite{AAK}, 
$L_{\varphi}$ does depend on temperature ($\propto T^{-1/3}$), 
while $L_{\rm SO}$ is almost constant at low temperatures~\cite{note_Debye}. 
Based on this fact, we fix $L_{\rm SO}$ at both temperatures to 
fit the WAL curves. 
We obtain $L_{\varphi}=4.20$ and 1.95~$\mu$m at $T=0.4$ 
and 4~K, respectively, 
while $L_{\rm SO}=800$~nm~\cite{pierre_prb_2003}. 
The two $L_{\varphi}$ values meet the Fermi liquid theory 
$L_{\varphi} \propto T^{-1/3}$. 
We have measured 4 different Ag wires on the same batch and 
obtained $L_{\rm SO} = 760 \pm 50$~nm. 

%As mentioned in the introduction, 
$L_{\rm SO}$ should be closely related to $L_{\rm s}$. 
This relation has been theoretically discussed in Ref.~\cite{zutic_rmp_2004}. 
The SO scattering rate
$1/\tau_{\rm SO} = D/L_{\rm SO}^{2}$ %~\cite{footnote} 
includes both spin-flip and spin-conserving processes, 
resulting in $1/\tau_{\rm SO} = 3/(2 \tau_{\uparrow\downarrow})$ where 
$1/\tau_{\uparrow\downarrow}$ is the spin-flip scattering rate.
We also note that the spin relaxation rate $1/\tau_{\rm s} = D/L_{\rm s}^{2}$ 
is twice the spin-flip scattering rate, i.e., 
$1/\tau_{\rm s} = 1/\tau_{\uparrow\downarrow} + 1/\tau_{\downarrow\uparrow}$. 
At sufficiently low temperatures, the contribution of phonons can be neglected 
and one obtains 
\begin{eqnarray} L_{\rm s} = \frac{\sqrt{3}}{2}L_{\rm SO}, 
\label{eq2} 
\end{eqnarray} 
within the EY mechanism from isotropic impurity scattering. 
Since Ag is a monovalent metal with an almost spherical Fermi surface, 
one can adapt Eq.~(\ref{eq2}) to convert from $L_{\rm SO}$ 
to $L_{\rm s}$.
We thus obtain $L_{\rm s} = 650 \pm 40$~nm, 
which is quantitatively consistent with 
$L_{\rm s} (\sim 600~{\rm nm})$ 
obtained from the lateral spin valve 
measurements~\cite{fukuma_apl_2010,hoffmann_prl_2010}. 
%{\color{red}
%Although one needs some ideal conditions for Eq.~(\ref{eq2}), 
%this is the first experimental
%verification of the relation between $L_{\rm SO}$ 
%and $L_{\rm s}$.
%}

%As is explained in Ref.~\cite{zutic_rmp_2004}, 
%\textit{``the SO scattering induces both spin-flip and 
%spin-conserving processes, which are in the ratio 2:1 
%for isotropic scattering rates. In addition, the 
%spin relaxation rate is twice the spin-flip scattering rate, 
%since each spin flip equilibrates both spins equally."}
%Therefore for isotropic systems one obtains 
%\begin{eqnarray}
%L_{\rm s} = \frac{\sqrt{3}}{2}L_{\rm SO}. \label{eq2}
%\end{eqnarray}
%Ag is a monovalent metal with an almost spherical Fermi surface. 
%Using Eq.~(\ref{eq2}) we obtain $L_{\rm s} = 650 \pm 40$~nm.
%This value is quantitatively consistent with 
%$L_{\rm s} (\sim 600~{\rm nm})$ obtained 
%from the lateral spin valve 
%measurements~\cite{fukuma_apl_2010,hoffmann_prl_2010}. 
%Thus, we conclude that $L_{\rm SO}$ is equivalent to $L_{\rm s}$ and 
%hereafter to simplfy the notation, we use only $L_{\rm sf}$ and focus 
%only on $L_{\rm sf}$ (not $L_{\varphi}$).

Next we discuss the SDL of a strong SO material such as Pt which is 
the most standard SHE material. 
As mentioned in the introduction, 
this is one of the causes of the big debates, i.e., 
$L_{\rm s} = 11$~nm from the spin absorption 
measurements~\cite{morota_prb_2011} and $L_{\rm s} = 1.4$~nm from 
the spin torque FMR measurements~\cite{liu_arXiv_2011}. 
To solve the problem,
here we perform two different measurements to obtain 
$L_{\rm SO}$ or $L_{\rm s}$ of Pt; 
WAL and spin absorption in the lateral spin valve devices. 
Note that the Pt wires for the WAL and spin absorption measurements 
were prepared at the same time. 
Figure~2(a) shows a typical WAL curve of Pt. 
The field scale is 20 times larger than that for 
Ag wires, which indicates that $L_{\rm SO}$ is much shorter. 
We observe maxima of the WAL curve at around $\pm 0.9$~T. %~\cite{note_WL_Pt}. 
From the best fit of Eq.~(\ref{eq1}), we obtain $L_{\rm SO} = 12$~nm. 
We have performed similar measurements for 5 different samples 
and using Eq.~(\ref{eq2}) we have determined 
$L_{\rm s}$ of Pt to be $10 \pm 2$~nm~\cite{note_Pt}. 

\begin{figure}
\begin{center}
\includegraphics[width=6.5cm]{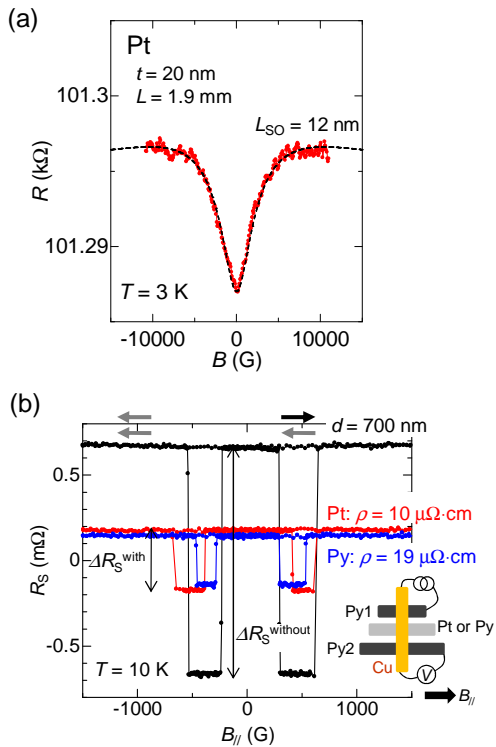}
\caption{(Color online) (a) WAL curve of Pt wire measured at $T=3$~K. The broken line is the best fit of Eq.~(\ref{eq1}). (b) NLSV signals with Pt ($R_{\rm S}^{\rm with}$; red) and Py (blue) wires in between the two Py wires. As a reference signal, we plot the NLSV signal without the middle wire ($R_{\rm S}^{\rm without}$; black). The width and thickness of the Pt or Py middle wire are 100~nm and 20~nm, respectively. The other dimensions are the same as in Refs.~\cite{niimi_prl_2011} and \cite{niimi_prl_2012}. A pair of arrows on the top indicates the magnetizations of Py1 and Py2. The inset shows the schemetic of our lateral spin valve device.} \label{fig2}
\end{center}
\end{figure}

The spin absorption measurement into the Pt wire is shown in Fig.~2(b). 
For comparison, Py middle-wire devices were prepared since 
the SDL of Py is well-known from other experiments~\cite{dubois_prb_1999}.
We have performed nonlocal spin valve (NLSV) measurements 
with and without the middle 
wires~\cite{niimi_prl_2011,niimi_prl_2012,morota_prb_2011}. 
The in-plane magnetic field $B_{\parallel}$ is applied 
parallel to the two Py wires [see the inset of Fig.~2(b)]. 
A pure spin current generated from Py1 is absorbed perpendicularly 
into the middle wire because of the strong SO interaction of Pt (or Py). 
As shown in Fig.~2(b), the NLSV signal detected at Py2 is reduced 
by inserting the Pt or Py wire compared to the one without any middle wire. 
To extract the SDLs of Pt and Py, 
we first use the 1D analytical model based on the 
Takahashi-Maekawa formula~\cite{takahashi_review_2008}. 
In this model, the normalized NLSV signal 
$\Delta R_{\rm S}^{\rm with}/\Delta R_{\rm S}^{\rm without}$ 
can be expressed as follows~\cite{niimi_prl_2011,niimi_prl_2012}; 
\begin{equation}
\frac{\Delta R_{\rm S}^{\rm with}}{\Delta R_{\rm S}^{\rm without}} 
\approx 
\frac{2R_{\rm M} \sinh (d/L_{\rm s}^{\rm Cu})}
{R_{\rm Cu} \left\{ \cosh (d/L_{\rm s}^{\rm Cu})-1 \right\} + 2R_{\rm M} \sinh(d/L_{\rm s}^{\rm Cu})} \label{eq3}
\end{equation}
where $R_{\rm Cu}$ and $R_{\rm M}$ are the spin resistances 
of Cu and the middle wire (Pt or Py), respectively.   
The spin resistance $R_{\rm X}$ of material ``X" is defined as 
$\rho_{\rm X} L_{\rm s}^{\rm X}/(1-p_{\rm X}^{2})A_{\rm X}$, 
where $\rho_{\rm X}$, $L_{\rm s}^{\rm X}$, $p_{\rm X}$ and $A_{\rm X}$ 
are respectively the electrical resistivity, the SDL, 
the spin polarization, and the effective cross sectional area 
involved in the equations of the 1D spin diffusion 
model~\cite{niimi_prl_2011,niimi_prl_2012}. 
$d$ is the distance between the two Py wires, in the present case $d=700$~nm. 
Although the spin absoprtion rate 
$\Delta R_{\rm S}^{\rm with}/\Delta R_{\rm S}^{\rm without}$ is 
almost the same for the Pt and Py middle wires, 
the obtained $L_{\rm s}$ from Eq.~(\ref{eq3}) 
are $11 (\pm 2)$~nm for Pt and $5 (\pm 1)$~nm for Py. 
This is because the resistivity of Pt is nearly half of Py. 
The SDL of Pt coincides well with that from our WAL measurement 
and the SDL of Py is also consistent with 
other experimental results~\cite{dubois_prb_1999}. 

We have also used the three-dimensional (3D) spin diffusion model 
based on the Valet-Fert formalism~\cite{valet_fert_prb_1993} 
to obtain $L_{\rm s}$ of Pt. 
This has been done in order to refute the claim made by Liu \textit{et al.} 
that $L_{\rm s}$ of Pt extracted from the 1D model 
might be overestimated~\cite{liu_arXiv_2011}. 
As detailed in Ref.~\cite{niimi_prl_2012}, 
SDLs obtained from the two methods do not 
differ significantly when the SDLs are comparable or smaller than 
the thickness of the middle wire. 
In fact, we have confirmed that the 3D analysis for the Pt middle wire
gives almost the same value as the 1D model. 
We thus conclude that the SDL of Pt itself is 
\textit{not} of the order of 1~nm \textit{but} about 10~nm. 

The reason why $L_{\rm s}$ of Pt reported in Ref.~\cite{liu_arXiv_2011} 
is much shorter than ours is that in the FMR measurement, 
the ferromagnet/nonmagnet bilayer is always used. 
In such a bilayer system, one cannot avoid the contribution from the magnetic 
damping effect~\cite{liu_arXiv_2011,mizukami_prb_2002,hillebrands_prl_2011}. 
As a result, the real SDL of nonmagnet can be modulated by the FMR, 
which results in a much shorter SDL. 
Thus, such a shorter SDL \textit{cannot} be adapted to 
the case of the spin aborption method and the SH angle of Pt 
should be about a few percent~\cite{morota_prb_2011}, \textit{should not} be 
enhanced up to 7\% as claimed 
by Liu \textit{et al}~\cite{liu_arXiv_2011}. 
Recently, Kondou \textit{et al.}~\cite{kondou_apex_2012} 
measured the SH angle of Pt using the same method as 
Liu \textit{et al.}~\cite{liu_prl_2011,liu_prl_2012,liu_arXiv_2011,liu_science_2012} 
but they carefully studied the thickness dependence of 
ferromagnet and nonmagnet. 
They found that the symmetric part of FMR spectra, from which 
the SH angle is extracted, does depend on the thickness of ferromagnet and 
one should take the zero-limit to avoid any effects 
from the ferromagnet~\cite{kondou_apex_2012}. 
The extrapolated SH angle is 0.022 for Pt, which 
is quantitatively consistent with that 
in Ref.~\cite{morota_prb_2011}.

\begin{figure}
\begin{center}
\includegraphics[width=8cm]{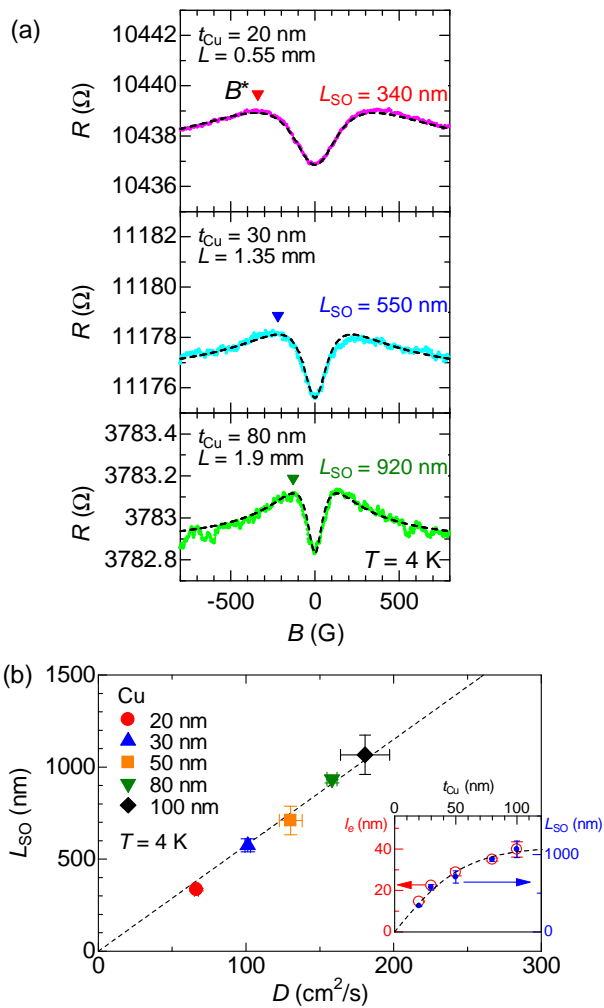}
\caption{(Color online) (a) WAL curves of Cu wires with different thicknesses ($t_{\rm Cu}=20$,~$30$, and $80$~nm) measured at $T=4$~K. The broken lines are the best fits of Eq.~(\ref{eq1}). The triangle in the figure corresponds to $B^{*}$. (b) Diffusion coefficient $D$ dependence of $L_{\rm SO}$ of Cu measured at $T=4$~K. The inset shows $l_{e}$ (left) and $L_{\rm SO}$ (right) as a function of $t_{\rm Cu}$. The broken line is a guide to eyes.} \label{fig3}
\end{center}
\end{figure}

%\begin{figure}
%\begin{center}
%\includegraphics[width=8cm]{Figure4.eps}
%\caption{(Color online) Diffusion coefficient $D$ dependence of the SDL $L_{\rm s}$ of Cu measured at $T=4$~K. The inset shows the elastic mean free path $l_{e}$ (left) and $L_{\rm s}$ (right) as a function of the Cu thickness $t_{\rm Cu}$. The broken line is a guide to eyes.} \label{fig4}
%\end{center}
%\end{figure}

To further support the comparability 
between $L_{\rm SO}$ and $L_{\rm s}$, we study 
the disorder effect on $L_{\rm SO}$. 
In metallic systems where $l_{e} < L_{\rm s}$, 
the EY mechanism is dominant for the spin relaxation process. 
If Eq.~(\ref{eq2}) is valid in metallic systems, 
$L_{\rm SO}$ should also follows the EY mechanism. For this purpose, 
we have chosen copper which has a weak SO interaction 
and then simply changed the thickness $t_{\rm Cu}$ of the copper wire. 
%As we will see later on, 
%we can change $D$ just by changing $t_{\rm Cu}$. 
Figure~3(a) shows WAL curves of 20, 30, 80~nm thick Cu wires. 
It is obvious that the maximum position (triangle in the figure) 
due to the SO interaction shifts toward lower fields with 
increasing $t_{\rm Cu}$. 
This position corresponds to the field where 
$2/L_{\rm SO} = w/l_{B}^{2}$, i.e., $B^{*}=2\hbar/(ewL_{\rm SO})$.
From the fitting, we obtain $L_{\rm SO}=340$, 550, 920~nm 
for $t_{\rm Cu}=20$, 30, 80~nm respectively. 

In Fig.~3(b) we plot $L_{\rm SO}$ of Cu as a function of $D$. 
Note that $D$ is determined from the Einstein relation $D = 1/(e^{2} \rho N)$ 
where $N$ is the density of state at the Fermi level~\cite{DOS_Cu}. 
$L_{\rm SO}$ nicely follows a linear law down to 20~nm thick Cu wires. 
According to the EY mechanism, $\tau_{\rm s}$ consists of the phonon 
and impurity contributions as follows; 
$1/\tau_{\rm s} = 1/\tau_{\rm s}^{\rm imp} + 1/\tau_{\rm s}^{\rm ph}$.
Since we focus on the low temperature part, 
we can neglect the phonon contribution~\cite{note_Debye} 
and concentrate on the discussion only about 
the spin relaxation from impurities, which makes 
the analysis much simpler~\cite{idzuchi_apl_2012}.
In addition, the impurity contribution can be expressed as 
$\tau_{\rm s}^{\rm imp} = \tau_{e}/ \varepsilon_{\rm imp}$ 
where $\tau_{e}$ and $\varepsilon_{\rm imp}$ are 
the elestic scattering time and 
the probability of spin-flip scattering, respectively~\cite{jedema_prb_2003}. 
Thus, one obtains the following equation; 
\begin{eqnarray}
L_{\rm SO}=\frac{2}{\sqrt{3}}L_{\rm s}=\frac{2D}{v_{\rm F}\sqrt{\varepsilon_{\rm imp}}} = \frac{2l_{e}}{3\sqrt{\varepsilon_{\rm imp}}} \label{eq5}
\end{eqnarray}
where $v_{\rm F}$ is the Fermi velocity. 
As can be seen in Fig.~3(b), the simplified EY mechanism Eq.~(\ref{eq5}) 
works down to at least $D \sim 60$~cm$^{2}/s$. 
Although there are many experimental works to determine the 
SDL~\cite{fukuma_apl_2010,takahashi_apl_2012,jedema_prb_2003} in the lateral spin valve structure, 
the disorder dependence of the SDL has not been focused on 
in those reports. 
As far as we know, there is only one report by Bass and Pratt 
to mention the disorder effect on $L_{\rm SO}$~\cite{bass_review_2007}. 
However, since they collected data from several different papers 
measured at different temperatures, 
it is not trivial to exclude the phonon contribution. 
%and the error bar is relatively large. 
%These facts are not allowed to verify Eq.~(\ref{eq5}).
Therefore, the present work is a clear experimental 
demonstration to verify Eq.~(\ref{eq5}).
From the fitting of $L_{\rm SO}$ vs $D$ curve, we estimate 
$\varepsilon_{\rm imp}$ to be $4.9 \times 10^{-4}$. 
This value is consistent with the ones obtained from 
the spin valve mesurement~\cite{jedema_prb_2003} and 
the CESR measurements~\cite{beuneu_prb_1976}.
We also show $L_{\rm SO}$ and $l_{e}$ 
as a function of $t_{\rm Cu}$ on the same plot in the inset of Fig.~3(b). 
Both $L_{\rm SO}$ and $l_{e}$ follow the same dependency.
When $t_{\rm Cu}$ is much larger than $l_{e}$, 
the impurity and defect contributions come mainly 
from the inside of the wires. 
With decreasing $t_{\rm Cu}$, $l_{e}$ is limited by $t_{\rm Cu}$. 
This means that the present system is perfectly 
diffusive and there is no specular scattering 
from the surface~\cite{niimi_prb_2010}. 
In this case, the surface scattering can be regarded as a kind of 
impurity or defect, and thus Eq.~(\ref{eq5}) works down to 
our lowest $D$.

In conclusion, we have experimentally verified 
the comparability between the SO lengths and the SDLs of noble metals 
using the WAL and spin absorption methods. This comparability 
works not only for weak SO materials but also for 
strong SO materials such as Pt.
We have also studied the disorder effect on the SO lengths of Cu 
by changing the thickness of wires. 
The obtained SO length nicely follows a linear law as a function 
of $D$, which clearly verifies the EY mechanism in the present system.

We acknowledge helpful discussions with C. B\"{a}uerle, 
S. Maekawa, S. Takahashi, S. Kasai and K. Kondou. 
We would also like to thank Y. Iye and S. Katsumoto 
for the use of the lithography facilities. 
This work was supported by KAKENHI. 
%(Grant No. 22840012, 24740217, and 23244071).


\begin{thebibliography}{00}
\bibitem{note_Debye}
The Debye temperature of Ag (Cu) is about 220 (320)~K. In the present situation ($T \leq 4$~K), the phonon contribution to $L_{\rm SO}$ is negligibly small.
\bibitem{note_Pt}
Strictly speaking, the prefactor in Eq.~(\ref{eq2}) would not be correct for Pt since Pt is not a simple monovalent metal. However, it should not be so different from $\sqrt{3}/2$.
\end{thebibliography}
\end{document}